\documentclass{emulateapj}
\usepackage{natbib}
\usepackage{bm}
\usepackage{mathrsfs,amsmath}
\usepackage{graphicx}
\usepackage{epstopdf}
\usepackage[english]{babel}
\usepackage{xcolor}
\usepackage[colorlinks,linkcolor={blue},citecolor={blue},urlcolor={blue}]{hyperref}
\shorttitle{StarGO}
\shortauthors{Yuan et al.}

\begin{document}

\title{StarGO: A NEW METHOD TO IDENTIFY THE GALACTIC ORIGINS OF HALO STARS}

\author{Zhen Yuan\altaffilmark{1,2,3}, Jiang Chang\altaffilmark{4}, Projjwal Banerjee\altaffilmark{2,3}, Jiaxin Han\altaffilmark{5,2,3}, Xi Kang\altaffilmark{4}, M. C. Smith\altaffilmark{1}}

\altaffiltext{1}{Key Laboratory for Research in Galaxies and Cosmology, Shanghai Astronomical Observatory, Chinese Academy of Sciences, 80 Nandan Road, Shanghai 200030, China; sala.yuan@gmail.com}
\altaffiltext{2}{Department of Astronomy, Shanghai Key Laboratory for Particle Physics and Cosmology, Shanghai Jiao Tong University, Shanghai 200240, China}
\altaffiltext{3}{IFSA Collaborative Innovation Center, Shanghai Jiao Tong University, Shanghai 200240, China}
\altaffiltext{4}{Purple Mountain Observatory, the Partner Group of MPI f{\"u}r Astronomie, 2 West Beijing Road, Nanjing 210008, China; changjiang@pmo.ac.cn}
\altaffiltext{5}{Kavli IPMU (WPI), UTIAS, The University of Tokyo, Kashiwa, Chiba 277-8583, Japan}
\begin{abstract}
We develop a new method \textsc{StarGO} (Stars' Galactic Origin) to identify the galactic origins of halo stars using their kinematics. Our method is based on self-organizing map (SOM), which is one of the most popular unsupervised learning algorithms. \textsc{StarGO} combines SOM with a novel adaptive group identification algorithm with essentially no free parameters. In order to evaluate our model, we build a synthetic stellar halo from mergers of nine satellites in the Milky Way. We construct the mock catalogue by extracting a heliocentric volume of 10 kpc from our simulations and assigning expected observational uncertainties corresponding to bright stars from Gaia DR2 and LAMOST DR5. We compare the results from \textsc{StarGO} against that from a Friends-of-Friends (FoF) based method in the space of orbital energy and angular momentum. We show that \textsc{StarGO} is able to systematically identify more satellites and achieve higher number fraction of identified stars for most of the satellites within the extracted volume. When applied to data from Gaia DR2, \textsc{StarGO} will enable us to reveal the origins of the inner stellar halo in unprecedented detail.

\end{abstract}


\section{Introduction}
According to the hierarchical structure formation theory, the Milky Way (MW) grows to its current size through frequent accretion and merger events. During these violent processes, satellite galaxies are tidally disrupted and the disk gets heated. The stellar halo is built up at the same time, as a repository of stars from various origins \citep{bullock05,font06,lucia08, cooper10, deason16}. Due to the approximately dissipationless nature of stars, substructures in the stellar halo, such as the stellar debris from a satellite or groups of stars that originated from the Galactic disk, may retain the memory of their origins. The identification of these substructures is the first step towards unraveling the evolution history of the MW. A number of such substructures have been found in the last decade, adding strong support to the scenario of hierarchical structure formation. One famous example is the discovery of the Sagittarius dwarf galaxy \citep{Ibata94, Ibata95, Yanny00} and its tidal streams \citep{Mateo96, Ibata01, Maj03}, both of which are located in the stellar halo.

The current hierarchical structure formation paradigm implies that the inner stellar halo contains a wealth of information about the early assembly history of the MW as the stars there tend to be accreted a long time ago. However, identifying substructures in configuration space is not easy due to the fact that the accreted substructures in the inner stellar halo have undergone mixing for a long time. Furthermore, this region is also populated by star groups likely originated from the disk, e.g. Monoceros \citep{ bergemann,laporte}, which makes substructure identification from satellites difficult.

On the other hand, identifying substructures in phase space can be relatively easier given the additional information from the velocities. In particular, the separations of stars in the integral-of-motion space are much better conserved and thus provide a natural coordinate system for identifying the original grouping of stars \citep{Helmi99, Helmi06, Smith09, Klement09, smith16r}. Previous searches of substructures in the inner stellar halo were hindered by the limited astrometric data. With the advent of \emph{Gaia}, we now have 5-D astrometric data for unprecedented number of stars (1332 million) from Data Release 2 \citep{dr2}. Cross matching TGAS \citep{TGAS} with other surveys such as RAVE \citep{RAVE}, LAMOST \citep{LAMOST}, 2MASS \citep{2mass}, and APOGEE \citep{APOGEE} has produced a stellar library within $\sim$ 20 kpc and has already led to several discoveries. For example, \citet{Koposov17} discovered faint MW satellites by searching for over-densities in configuration space, \citet{Helmi16} found a substructure of halo stars in integral-of-motion space, and \citet{Myeong17} identified the existence of a comoving star cluster with additional information of metallicity distribution.

Despite the increasing discovery of identified substructures, their number is far below the predictions from $\Lambda$CDM cosmology. According to Aquarius simulations \citep{aquarius}, hundreds of streams are expected in the $Gaia$ sky \citep{gomez13, maffione15}. In order to systematically identify these substructures using the vast amount of astrometric data, several methods have been developed, including distance based methods such as Friends-of-Friends (FoF) \citep{Helmi00}, and density based algorithms such as Mean Shift\citep{gomez10a} and Watershed\citep{Helmi16}. 

In this paper, we propose a new method of substructure identification, that requires essentially no free parameters. Our method utilizes a machine learning technique called self-organizing map (SOM) \citep{kohonen82}, that maps out the topology of a high-dimensional dataset onto a two-dimension map. Using the fact that stars with the same origin have similar orbital energy and angular momentum, we first apply SOM to the n-Dimensional (n-D) input space constructed from these quantities and visualize the results in a 2D neural map. Then, we develop a new adaptive group identification scheme based on the resulting 2D map. Since SOM retains the topological structure of the data set, it can manifest the fine structures in the data. This makes our method particularly well suited in identifying groups that are weakly clustered. We test the performance of our method by applying it to a mock catalogue generated from our simulation of a MW-like system with realistic observational uncertainties. 

The paper is organized as follows; Sec.~\ref{sec:simu}: details of model setup and simulations for generating the mock catalogue, Sec.~\ref{sec:method}: details of SOM and group identification of \textsc{StarGO}, Sec.~\ref{sec:res}: results of \textsc{StarGO} applied to the mock catalogue and comparisons with FoF, Sec.~\ref{sec:sum}: conclusion. 

\section{Simulations}\label{sec:simu}
\subsection{Overview}\label{subsec:overview}
A popular approach of building a stellar halo is using zoom-in $\Lambda$CDM cosmological simulation of a MW-like system accompanied by post-process of star tagging using semi-analytic models such as \textsc{galform} \citep{lucia08,cooper10,tumlinson10}. Although such models can retain the realistic accretion history of a MW-like system, they do not include any stellar components such as the disk and the bulge, which are crucial for modeling the kinematics of stars in the inner stellar halo.

Another approach involves using an analytic potential for the dark matter and stellar components of the MW, while using N-body models for the dark matter component of satellites \citep{Helmi00}. The main advantage of this approach is that it can achieve higher resolution than pure N-body simulations. Instead of using the actual merger history, an artificial one is used to build up a synthetic stellar halo. \citet{gomez10b} used a similar approach but with a time dependent MW potential. In such studies, stars in satellites are assigned to particles in post-process and \textit{in situ} stars are added as background contamination. In a different approach of resimulation of a MW-like system, \citet{jb17} modeled the MW and satellites as collections of both dark matter and star particles to get a live N-body simulation.

In this study, we use a static analytic potential to model the dark matter halo of the MW, while particles are used to model stars in the disk and the bulge. For satellites, particles are used for both dark matter and stars. Although our method cannot account for dynamical friction since we use an analytic potential, it is expected to have a minor effect for the mass range of satellites chosen in our study \citep{frings17,amorisco17}. This makes our method computationally much less expensive compared to studies which use live models (e.g. \citealt{jb17}).

Similar to the second approach mentioned above, we use an artificial merger history. Zoom-in cosmological simulations of MW-like systems suggest that the main contributors to the inner stellar halo are a few satellites that infall at early times \citep{lucia08,cooper10}. In this study, we build up an synthetic stellar halo through several minor mergers following the same ideas as \citet{boylan08} and \citet{amorisco17}, where the dynamical friction from dark matter halo can be neglected. On the other hand, energy and angular momentum exchange between the MW center (the disk and the bulge) and satellites is important since all the satellites have peri-center distances within 20kpc (see Table~\ref{tab:orbit}). This however, is automatically taken into account as they are modeled using particles. Here, we model the accretion of 9 satellites which infall in the first 0 -- 4 Gyr (see Tab.~\ref{tab:orbit}), where the total simulation time is 12 Gyr. 

We use progenitors with infall mass ratios relative to the virial mass of MW of $\lesssim$ 1:25. In this mass range, even under dynamical friction from the halo, the initial orbital imprints of satellites can be retained in some of the stars \citep{amorisco17}. The infall radial velocity $V_r$ and tangential velocity $V_{\theta}$ are set to be some fraction of the virial velocity of the MW $V_{\mathrm{vir,MW}}$. The adopted value of the fractions are taken from the preferred range from cosmological simulations by \citet{jiang14}. We calculate the circularity $j$, defined as the ratio of the total angular momentum to the angular momentum for a circular orbit of the same energy. The satellites in our model have $j=$ 0.2 -- 0.8, which are consistent with values found in studies of accreted satellites from cosmological simulations \citep{jiang14} as well as from other models of the stellar halo \citep{amorisco17}.

\subsection{Our Model}
\label{subsec:mod}

The dark matter halo of the MW is described as a Navarro-Frenk-White (NFW) potential \citep{Nava} with virial mass $M_{vir}$ = $10^{12}$M$_{\odot}$ and concentration parameter c = 7 \citep{maccio08}. The stellar part of the MW consists of a Hernquist bulge \citep{Hern} and an exponential disk with the total stellar mass of M$_{\ast}$ = 0.03M$_{\rm vir}$ and particle mass of 3$\times10^4$M$_{\odot}$. The bulge component contributes $\sim$20$\%$ of M$_{\ast}$, with a density profile $\rho_{\rm b}$ given by
\begin{equation} \label{eq:hern}
\rho_{\rm b}(r)=\frac{M_{\rm b}}{2\pi}\frac{a}{r(r+a)^3},
\end{equation}
where $M_b$, $a$, and $r$ are the mass, scale length, and radius, respectively. The rest $\sim$80$\%$ of the total stellar mass is in the stellar disk of mass M$_d$ with the density profile parameterized by scale length $R_{\rm s}$ and scale height $z_0$ = 0.2$R_{\rm s}$, given by
\begin{equation} \label{eq:exp}
\rho_{\rm d}(r, z)=\frac{M_{\rm d}}{4\pi z_0 R_{\rm s}^2}\mathrm{sech}^2(\frac{z}{2z_0})\exp(-\frac{R}{R_{\rm s}}),
\end{equation}
where $z$ and $R$ are the height and radius, respectively.

For pre-cooked progenitors of satellites, we use similar recipe to \citet{chang}. Each satellite has a dark matter halo with NFW profile and an exponential disk with a total stellar mass of $1\%$ of the virial mass of the satellite. We use particle mass of $5\times10^4$M$_{\odot}$ and $5\times10^3$M$_{\odot}$ for the dark matter halo and stellar disk, respectively. We design three types of progenitors H-m, M-m, and L-m with total masses of 4$\times$10$^{10}$M$_{\odot}$, 10$^{10}$M$_{\odot}$ and 2.5$\times$10$^9$M$_{\odot}$, respectively (see Table~\ref{tab:icparam}).

\begin{table*}
\centering
\caption{Properties of the MW and satellite galaxies}\label{tab:icparam}
	\begin{tabular}{ccccccccc}
		\hline
    Type & $M_{\mathrm{vir}}$(M$_{\odot}$) & $V_{\mathrm{vir}}$(km/s)& $c$ & $N_h$ & $M_{\ast}$(M$_{\odot}$) & $R_s$(kpc/h) & $N_{\ast}$ & $m_{p,\ast}$(M$_{\odot}$) \\
  \hline
  MW & $10^{12}$          &162.6 & 7.0  &  -      & 3$\times10^{10}$ & 3.01 & 6$\times10^5$ & 5$\times10^4$ \\
  H-m & 4$\times10^{10}$  &55.6  & 9.3  & 7.84$\times10^5$ & $8\times10^8$    & 0.98 & 1.6$\times10^5$ & 5$\times10^3$ \\
  M-m & $10^{10}$         &35.0  & 10.6 & 1.96$\times10^5$ & $2\times10^8$    & 0.57 & 4$\times10^4$  & 5$\times10^3$  \\
  L-m & $2.5\times10^{9}$ &22.1  & 12.2 & 4.9$\times10^4$  & $5\times10^7$    & 0.34 & 10$^4$  & 5$\times10^3$  \\
  \hline
	\end{tabular}
  \medskip

For each type of galaxy, $M_{\mathrm{vir}}$ and $V_{\mathrm{vir}}$ are the viral mass and viral velocity (for satellites these values refer to their initial values), respectively. Note that $R_{\rm vir}$ has the same value as $V_{\rm vir}$, with the unit of kpc/h. For the dark matter halo part: $c$ is the concentration parameter of the NFW model, $N_{\rm h}$ is the number of particles in the dark matter halo, and $m_{\rm p,h}=$5$\times$10$^4$M$_{\odot}$ is the mass for each dark matter particle. For the stellar part: $M_{\ast}$ is the stellar mass, $R_{\rm s}$ is the disk scale length, $N_{\ast}$ is the number of star particles, and $m_{\rm p,\ast}$ is the mass for each star particle.

\end{table*}

\begin{figure*}[tb]
\centering
\includegraphics[width=\linewidth]{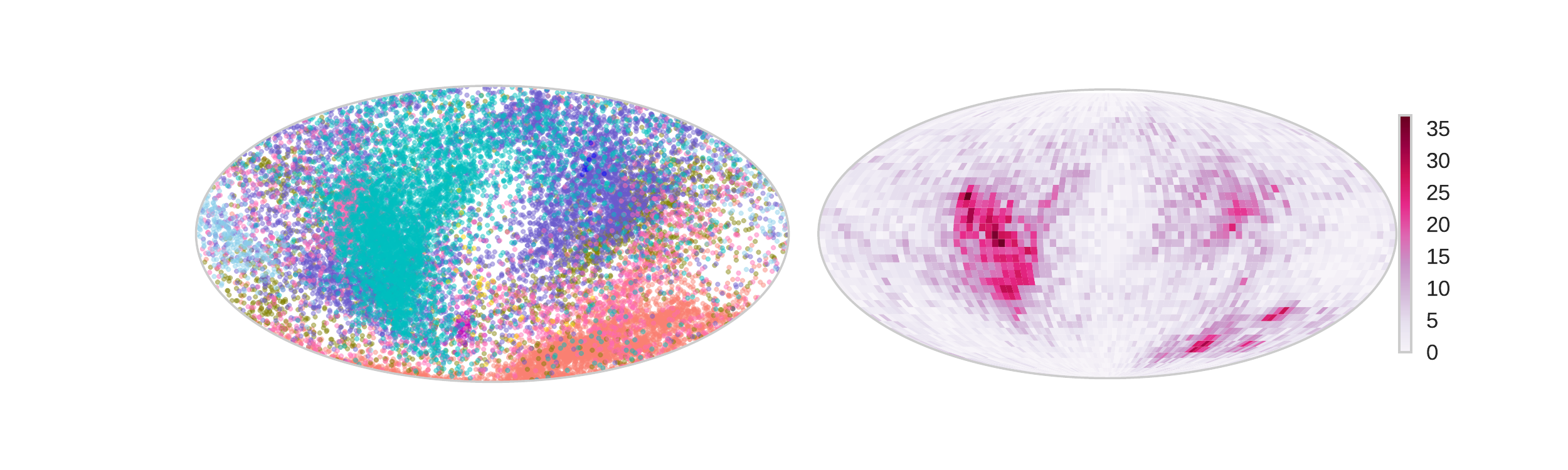}
\caption{Aitoff projection of angular momentum of stars from mock catalogue in the galactic reference frame. Left panel: Stars from different satellites are assigned unique colors as follows: sat1 (salmon), sat2 (olive), sat3 (magenta), sat4 (purple), sat5 (pink), sat6 (blue), sat7 (cyan), sat8(gold), sat9(light blue). Right panel: The corresponding density map of all the satellite stars in the same projected space as the left panel. }
\label{fig:proj_L}
\end{figure*}
\begin{figure*}[tb]
\centering
\includegraphics[width=\linewidth]{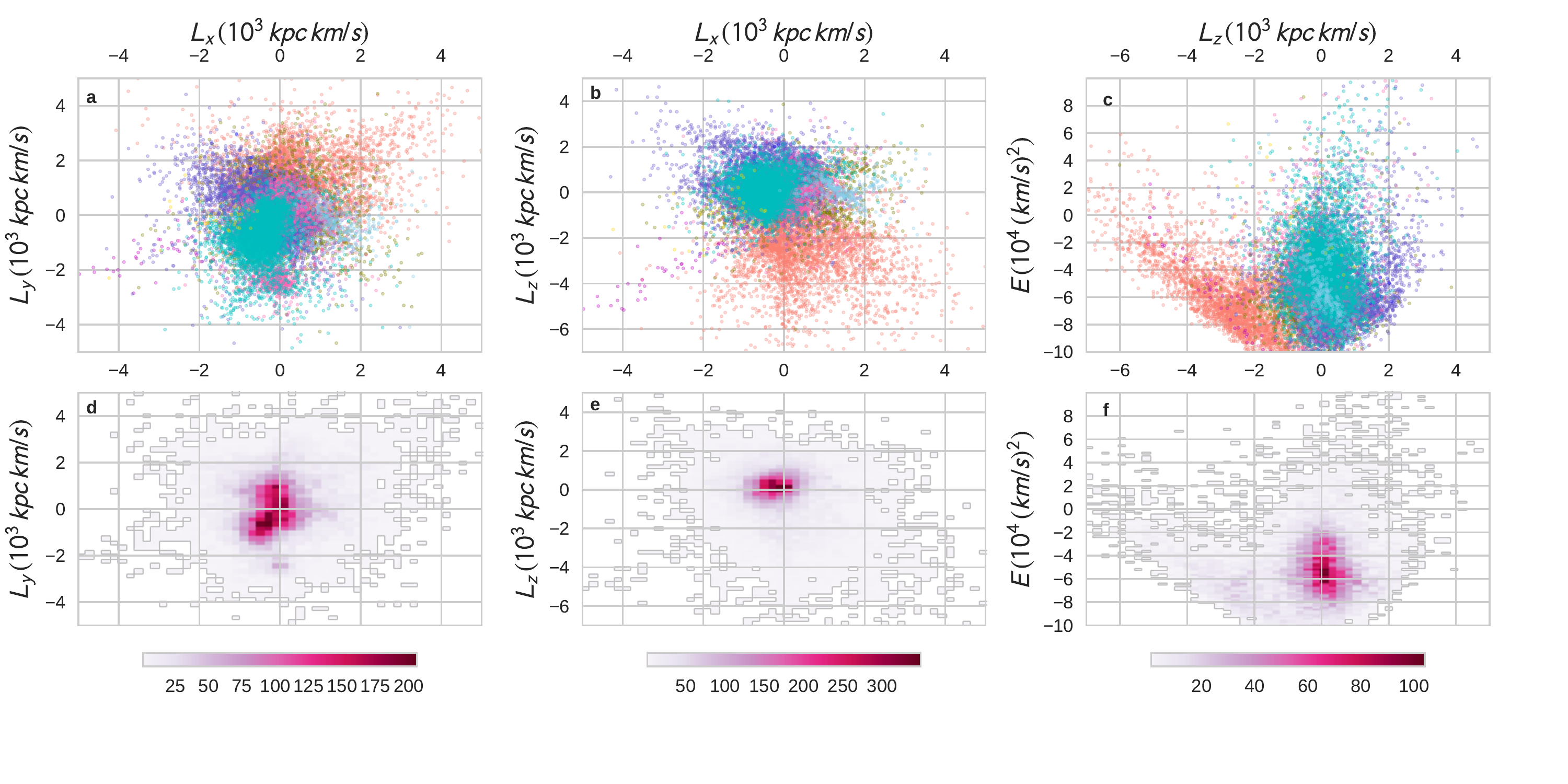}
\caption{Top row: Projection of all the stars from the mock catalogue in (a) $L_x$ - $L_y$, (b) $L_x$ - $L_z$, and (c) $L_z$ - $E$ plane, using the same color codes of Fig.~\ref{fig:proj_L}. Bottom row (b) (c) (d): The corresponding density map of all the stars in the same plane as the top row.}
\label{fig:proj_EL}
\end{figure*}
Using H-m, M-m, and L-m, we create nine satellites (sat1 -- 9) with distinct infall scenarios by changing their initial velocities and positions (see Tab.~\ref{tab:orbit}). All the satellites are released at distance $R_{\mathrm{ini}}$ with an initial radial velocity $V_{\rm r}$ and tangential velocity $V_{\rm \theta}$ \citep{benson05,jiang14}. The inclination of the satellite orbit with respect to the disk is characterized by $i$, which is the angle between initial orbital direction of the satellite and the initial Galactic $z$ direction. The values of $i$ are chosen from a range of 0--60$^{\circ}$. We choose $R_{\mathrm{ini}}$ to be less than the virial radius for all the satellites to emulate the early infall scenarios when the MW is smaller.

\begin{table*}
\centering
\caption{Details of the mock catalogue}\label{tab:orbit}
	\begin{tabular}{ccccccccc}
		\hline
		     & Type & $T_{\mathrm{inf}}$(Gyr) & $R_{\mathrm{ini}} / R_{\mathrm{vir,MW}}$ & $(V_{\rm r}, V_{\rm \theta}) / V_{\mathrm{vir,MW}}$ & $i(^{\circ})$ &$r_{\mathrm{peri}}$(kpc) & $j=J/J_{\mathrm{circ}}$& $N_{\mathrm{sat}}$ \\
  \hline
  sat1   & H-m  & 0 & 0.4  & ( 0.64, -0.64) & 60 & 22.8 & 0.71 & 3,609     \\
  sat2   & M-m  & 0 & 0.4  & ( 0.96,  0.32) & 45 &  9.1 & 0.32 & 1,970     \\
  sat3   & L-m  & 0 & 0.4  & ( 1.0,   0.2)  & 30 &  5.4 & 0.20 & 59     \\
  sat4   & H-m  & 2 & 0.6  & ( 0.96,  0.32) & 45 & 13.2 & 0.32 & 5,598   \\
  sat5   & M-m  & 2 & 0.6  & ( 0.96, -0.32) & 0 & 13.2 & 0.32 & 1,961     \\
  sat6   & L-m  & 2 & 0.6  & ( 0.72,  0.72) & 45 & 38.7 & 0.71 & 9     \\
  sat7   & H-m  & 4 & 0.8  & ( 0.6,  -0.2)  & 0 & 10.6 & 0.32 & 4,985     \\
  sat8   & M-m  & 4 & 0.8  & ( 0.36,  0.36) & 30 & 22.2 & 0.71 & 38     \\
  sat9   & L-m  & 4 & 0.8  & ( 0.48,  0.16) & 15 &  8.4 & 0.32 & 341    \\

  \hline
	\end{tabular}
  \medskip

The initial condition of each satellite before it falls into the MW at $T_{\mathrm{inf}}$ are represented by the distance $R_{\mathrm{ini}}$, velocity $(V_{\rm r}, V_{\rm \theta})$, and inclination angle $i$. The peri-center distance and circularity of the orbit for each satellite are denoted as $r_{\mathrm{peri}}$ and $j$, respectively. The number of stars from each satellite $N_{\mathrm{sat}}$ is shown in the last column.
\end{table*}

\subsection{Catalogue} \label{subsec:cat}
In order to generate data sample from our simulation, we select all the stars within 10 kpc relative to the Sun, which is taken to be at the galactocentric distance $R_{\odot}=8$ kpc. This is done to get data with the coverage similar to the $Gaia$ sky. We address the sampling bias introduced by the solar position by generating 8 samples, where the solar position is rotated by 45$^{\circ}$ each time in the x-y plane. We find the samples to be quantitatively similar, i.e. number of stars in each population is similar. In this paper, we use only one of the samples for demonstration, the details of which are given in Table~\ref{tab:orbit}.

The disk and the bulge population accounts for more than 95$\%$ in our sample, most of which are distributed close to the galactic disk plane. Since we are mainly interested in the halo stars, we exclude all the stars from these two components to generate the mock catalouge. In real observational samples, this can also be done relatively easily by using a cut for metallicity or distance from the mid-plane. The total number of stars in each satellite is denoted by $N_{\mathrm{sat}}$ (see Tab.~\ref{tab:orbit}). Most of the stars in the extracted heliocentric volume come from six of the nine satellites (sat1, sat2, sat4, sat5, sat7, and sat9).

K-giants are ideal tracers of the stellar halo because they are bright and distances can be reliably estimated from photometry. Therefore we construct a mock catalog of K-giant stars, adopting errors similar to what will be obtained for a sample of such stars from LAMOST and Gaia DR2. Specifically, we assign the distance error to be 20\% according to the distance estimation method using photometry \citep{xue14}.  We adopt the error of radial velocity to be 7 km/s which is consistent with LAMOST  \citep{sch17}. The proper motion error at G$\sim$16--17 mag is 0.1--0.2 mas/yr from Gaia DR2 \citep{dr2}. We take 0.15 mas/yr for all the stars in the catalogue. This is a reasonable number for K-giants, since even at 10 kpc, they have G$\sim$ 16--17 mag. The corresponding error in tangential velocity for a star at 10 kpc is about 7 km/s, i.e. comparable to our assumed error in radial velocity from low-resolution spectroscopy.

Before we discuss the details of our group identification method and the results, it is useful to first visualize the input data. We plot stars in the Aitoff projection of angular momentum space shown in Fig.~\ref{fig:proj_L}. The left panel shows stars from different satellites whereas the right panel shows the corresponding density map. We can clearly see that stars of the same origin tend to cluster, although stars of different origins often overlap each other. This can also be seen from Fig.~\ref{fig:proj_EL}a--c where stars are plotted in ($L_x$, $L_y$), ($L_x$, $L_z$), and ($L_z$, $E$). The clustering can be clearly seen in the corresponding density maps (see right panel of Fig.~\ref{fig:proj_L} and Fig.~\ref{fig:proj_EL}d--f). This clustering information is the basis of all substructure identification methods. Below we discuss the details of our method applied to different input spaces discussed above.

\section{Method} \label{sec:method}

Here, we develop a substructure identification method based on SOM, which belongs to unsupervised learning domain. We apply SOM to the mock data catalogue, followed by a novel group identification procedure which utilizes the visualization of SOM output in the 2D neural map. Below we give a brief introduction to SOM followed by details of our novel group identification procedure.

\begin{figure*}[tb]
\centering
\includegraphics[width=\linewidth]{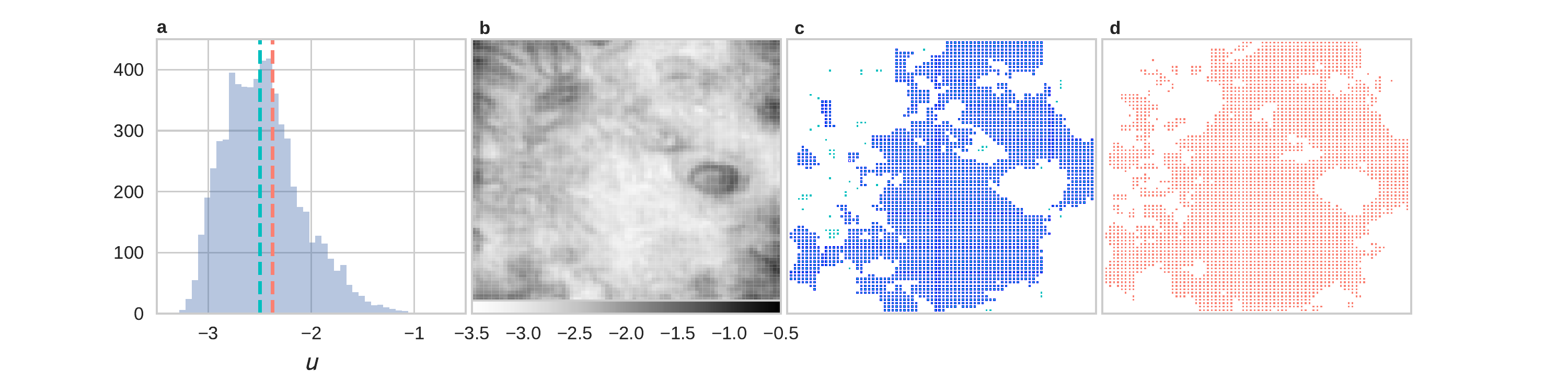}

\caption{Results from the first iteration of workflow of \textsc{StarGO} to the mock catalogue in the ($E$, $L_x$, $L_y$, $L_z$) space. a: The blue shaded histogram shows the distribution of $u$, where the cyan and salmon dashed lines denote $u_{\rm m}$ and $u_{\rm thr}$ values, respectively. b: 2-D neural map resulting from SOM, where the $u$ value between adjacent neurons is represented by the gray color scale. c: The same map as b which shows the selected neurons in different colors according to  step 2--3 of the workflow. 
The neurons with $u < u_{\rm m}$ are marked by colored pixels; each pixel associated to a group is colored cyan, while groups with more than 30 stars have their pixels enclosed with a blue box. d: The neurons with $u<u_{\rm thr}$ are marked by salmon pixels according to step 4 of the workflow. }
\label{fig:u}
\end{figure*}

\begin{figure}[tb]
\centering
\includegraphics[width=\linewidth]{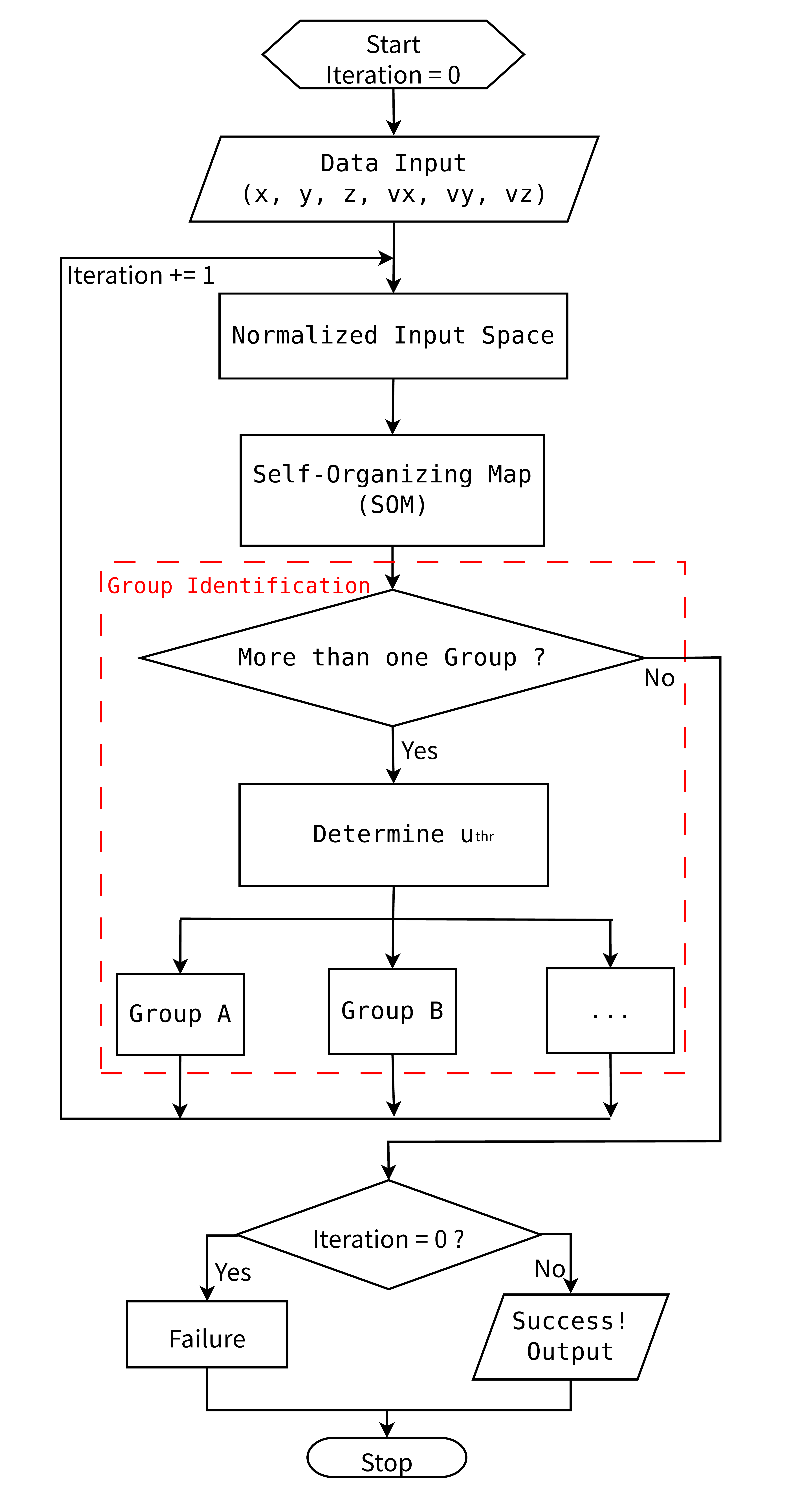}
\caption{The workflow of \textsc{StarGO}}
\label{fig:workflow}
\end{figure}

\begin{figure*}[tb]
\centering
\includegraphics[width=\linewidth]{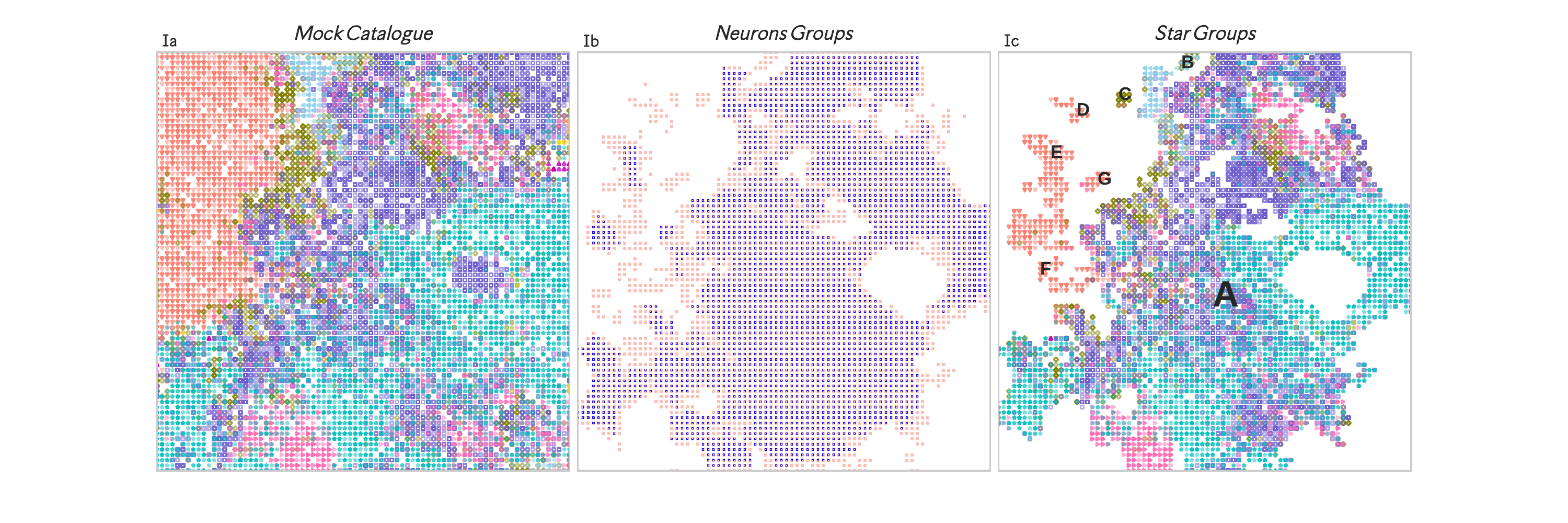}
\includegraphics[width=\linewidth]{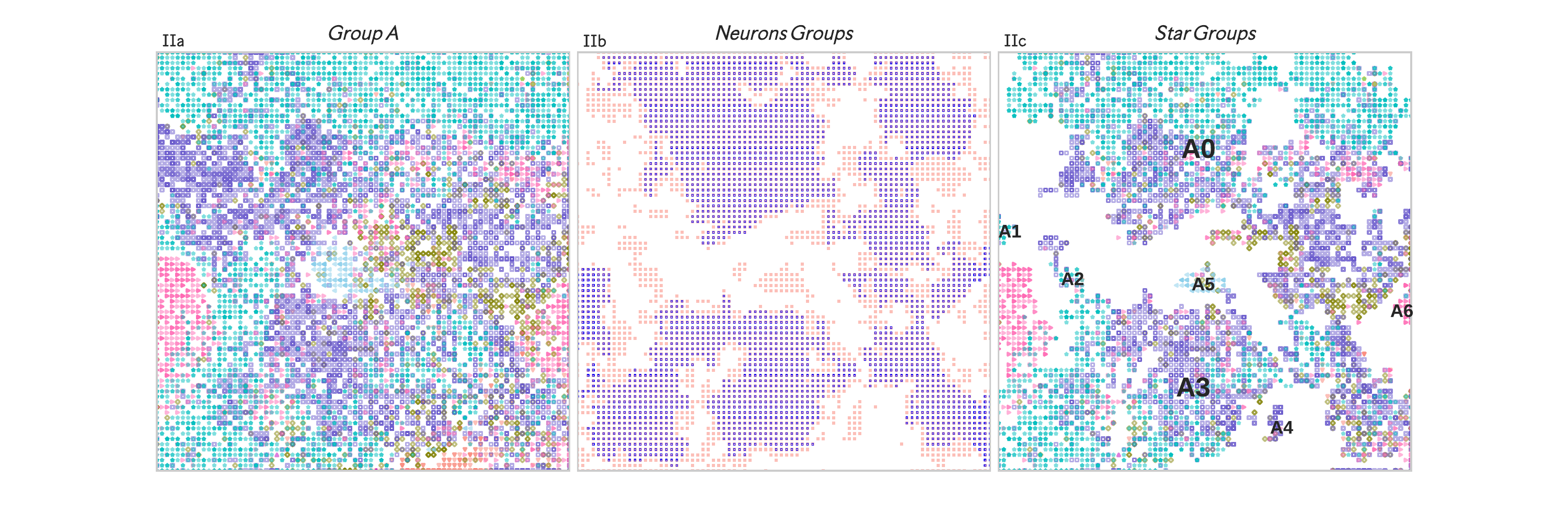}
\includegraphics[width=\linewidth]{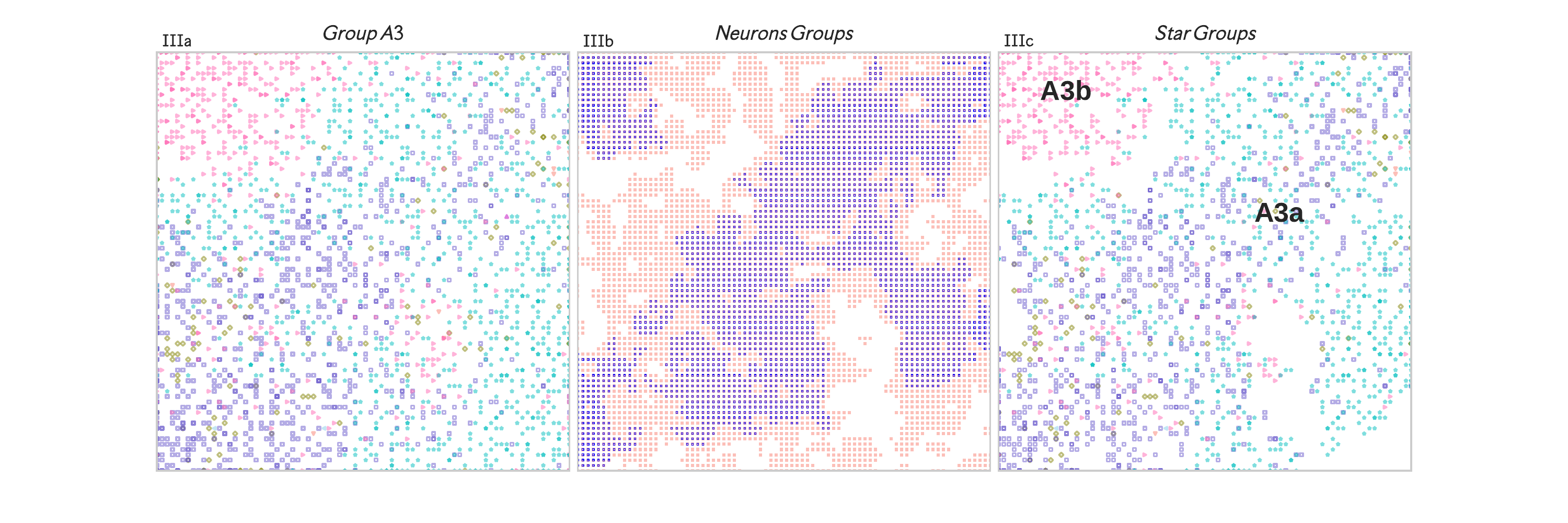}

\caption{Illustration of \textsc{StarGO} workflow applied to the mock catalogue in the ($E$, $L_x$, $L_y$, $L_z$) space for three iterations. Ia: Direct map of stars to their BMUs after the initial iteration of SOM. Stars in each satellite are assigned with unique color (same as Fig.~\ref{fig:proj_L}) and symbol: sat1 (\emph{salmon upper triangle}), sat2 (\emph{olive right triangle}), sat3 (\emph{magenta plus}), sat4 (\emph{purple diamond}), sat5 (\emph{pink cross}), sat6 (\emph{blue square}), sat7 (\emph{cyan star}), sat8 (\emph{gold hexgon}), and sat9 (\emph{light blue left triangle}). Ib: On the same map of Ia, the neurons in seed groups are marked by blue pixels, same as Fig.~\ref{fig:u}c. The neurons with $u < u_{\rm thr}$ are marked by salmon pixels, same as Fig.~\ref{fig:u}d. Ic:  On the same map of Ia, the identified star groups  (Group A--G) are plotted, where stars in each group are mapped to their BMU with the same color coding as Ia. IIa: Direct map of stars from Group A after the second iteration of SOM. IIb: On the same map of IIa, the selected neurons according to step 3--4 from the second iteration of the workflow are plotted. IIc:  On the same map of IIa, the identified star groups (Group A0--A6) are plotted. IIIa: Direct map of stars from Group A3 after the third iteration of SOM. IIIb: On the same map of IIIa, the selected neurons are plotted from the third iteration of the workflow. IIIc: On the same map of IIIa, the identified star groups (Group A3a--A3b) are plotted.}
\label{fig:stargo}
\end{figure*}

\begin{figure}[tb]
\centering
\includegraphics[width=\linewidth]{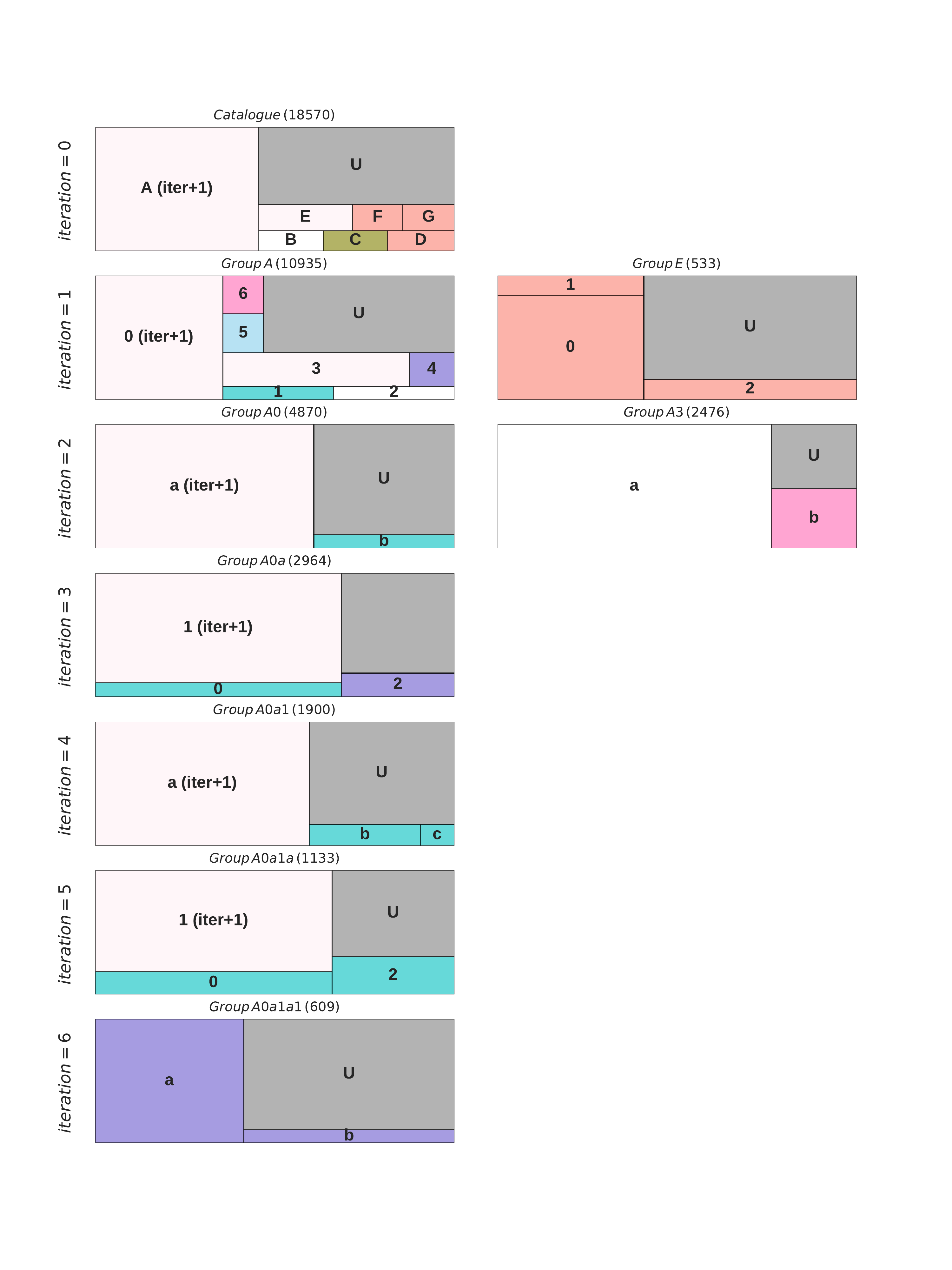}

\caption{Schematic diagram for the hierarchical group identification from \textsc{StarGO}. The results for star groups after each iteration is represented by tree maps in each row. Every tree map contains rectangles of different colors and areas, each of which represent a distinct star group. 
Indivisible groups with purity $\geq 60\%$ are colored according to the major contributor with the same color scheme as in Fig.~\ref{fig:proj_L}. If an indivisible group cannot be associated to any dominant single satellite, the rectangle is colored in white. The divisible groups are denoted by light pink rectangles, and the unidentified stars (i.e. stars not associated to any group) are denoted by grey rectangles. The relative area of each rectangle in a tree map approximately represents the number fraction of stars from the corresponding group. }
\label{fig:treemap}
\end{figure}

\subsection{Self-Organizing Map} \label{subsec:som}
The aim of SOM is to map a n-D input data to a 2D neural map while retaining the topological structures within the data at the same time. The starting point is the construction of a 2D map of $m\times m$ neurons, each of which are located at a different grid point $(a, b)$. Neurons have initially randomized weight vectors $\mathbf{w}$ with the same dimension and range as the n-D input vectors $\mathbf{v}$. Given an input vector $\mathbf{v}^i$, for the $i$-th star from the data catalogue, we first find the neuron that has the closest weight vector to $\mathbf{v}^i$ by finding the neuron with the minimum value of $|\mathbf{w} - \mathbf{v}^i|$. Such a neuron is defined to be the best matching unit (BMU). The learning process involves improving weight vectors $\mathbf{w}$ of all the neurons towards $\mathbf{v}^i$ according to their distances $d^{i}_{a,b}$ to the BMU located at ($a_i$, $b_i$) on the 2D neural map, where $d^{i}_{a,b}$ is defined as
\begin{equation}
d^{i}_{a,b} = \sqrt{(a - a_i)^2 + (b - b_i)^2}.
\end{equation}
The change in the weight vector $\mathbf{dw}^i_{a,b}$ due to the $i$th star is given by
\begin{equation}
\mathbf{dw}^i_{a,b} = \alpha_q\exp\left(-\frac{{d^{i}_{a,b}}^2}{\sigma^2_q}\right)(\mathbf{v}^i - \mathbf{w}^i_{a,b}),
\end{equation}
where $\alpha_q$ characterizes the learning rate and $\sigma_q$ controls the neighboring influence of neurons around the BMU for the $q$-th iteration. We can see from the above equation that the change in the weight of a neuron is sensitive to its distance from the BMU. The learning process is performed using $\mathbf{v}^i$ for each star in the data set. The learning process is then repeated for a total number of $N_{\rm iter}$ iterations. For the q$th$ iteration, the corresponding $\alpha_q$ and $\sigma_q$ is
\begin{equation}
\alpha_q = \alpha_0 (1 - q/N_{\rm iter}), \quad
\sigma_q = \sigma_0 (1 - q/N_{\rm iter}),
\end{equation}
where we use typical fiducial values of $\alpha_0 = 0.3$ and $\sigma_0$ = max($m$, $n$)/2 similar to \citet{som12}. We find that the results are independent of $\alpha_0$ and $\sigma_0$ for reasonable variation around this fiducial value. As the number of iteration $q$ increases, $|\mathbf{dw}|$ is reduced due to the decrease in $\alpha_q$ and $\sigma_q$, leading to refinement of the learning process. The learning process is considered to be complete when $|\mathbf{dw}|/|\mathbf{w}|\rightarrow$ 0.

\subsection{Group Identification}
\label{subsec:gi}

We feed the mock catalogue in a given input space to a 80$\times$80 neural network. After the application of SOM, the clustering structures can be visualized by using the differences between the weight vectors of neighboring neurons, which are the elements of the $u$-matrix defined as

\begin{equation}
u_{a,b} = \log_{10}\left(|\mathbf{w}_{a\pm1, b} - \mathbf{w}_{a, b}| + |\mathbf{w}_{a, b\pm1} - \mathbf{w}_{a, b}|\right).
\end{equation}

Neurons mapped to stars in highly-clustered regions tend to have similar angular momentum and orbital energy, leading to lower values of $u$ and vice versa. Fig.~\ref{fig:u}a shows the distribution of $u$ while Fig.~\ref{fig:u}b shows the resulting 2-D 80$\times$80 neural map, where $u$ is represented by the gray color scale. Each star can be mapped to its BMU in the 2-D map. We note that every neuron can be associated with more than one star or no stars at all.

Based on the 2-D map generated by SOM, we develop a novel algorithm for identification of substructures. Below, we list the steps adopted for group identification starting with the application of SOM followed by our algorithm:

\begin{enumerate}
  
  \item We first normalize each component of the input vector. For each dimension of the input vector we calculate the 95\% confidence interval for the whole sample. We then divide each component of the input vector for each star by this normalizing factor. We then apply SOM to all the stars in the normalized input space and calculate $u$-matrix from the resulting map. We associate each star to its BMU in the 2-D map.
  
  \item We group neighbouring neurons with $u<u_{\rm m}$ to form candidate seed groups (marked by cyan pixels in Fig.~\ref{fig:u}c), where $u_{\rm m}$ is the median value of u for all neurons in the map.

  \item A candidate seed group resulting from Step 2 is considered as a bona fide seed group (enclosed with blue boxes in Fig.~\ref{fig:u}c) if more than 30 stars are associated to neurons in the group.

 \item If there are more than one bona fide seed group, we maximize the size of each group by increasing the value of $u_{\rm thr}$ (shown as salmon dashed line in Fig.~\ref{fig:u}a). This results in the merging of multiple groups and we increase $u_{\rm thr}$ until we have two groups remaining from our original set of seed groups (marked by salmon pixels in  Fig.~\ref{fig:u} d). We stop increasing $u_{\rm thr}$ when these two groups are as large as possible, i.e. if $u_{\rm thr}$ was increased any further then these would merge. Additional groups can arise as the increased $u_{\rm thr}$ results in the formation of some new groups which have more than 30 stars associated to them.

  \item Star associated with the identified neuron groups form the corresponding identified star groups. For each identified group, we apply SOM followed by the above group identification procedure by repeating step 1 to 4 (see workflow in Fig.\ref{fig:workflow}). Stars not belonging to any group are designated as``unidentified".

  \item We stop the group identification procedure when no more than one seed group can be found after step 3.
  
\end{enumerate}

The group identification algorithm above allows us to find substructures adaptively for a given data set. At each iteration, neurons with $u<u_{\rm m}$ correspond to stars that have clustering above the median value. We set the minimum number of stars for seed groups to be 30 to discard small groups with low significance. The threshold of identified group is also set to be 30, which is slightly below the number of stars from the smallest population in the catalogue. Increasing the value of $u_{\rm thr}$ in step 4 is designed to maximize the completeness of grouped stars. Since we apply our algorithm iteratively to each group, the final set of groups represent the smallest indivisible group that has at least 30 stars.

\subsection{Size and Convergence Check}
\label{subsec:check}

We check the dependence of the results on the size of neural network, by using networks of size 50$\times$50, 80$\times$80, and 100$\times$100 neurons. We find that the convergence is achieved for network size of 80$\times$80, with the larger network yielding almost identical results. The smaller network fails to identify fine structures due to coarse griding. Thus, we use the network of size 80$\times$80 throughout the study. We perform additional checks on dependence of the results on the iteration number $N_{\rm iter}$. We use $N_{\rm iter}$ = 200, 300, and 400, finding the results are already converged for $N_{\rm iter}$ = 200. We adopt $N_{\rm iter}$ = 200 as the default value.

\section{Results and Discussions} \label{sec:res}

In this section, we apply \textsc{StarGO} to the mock catalogue in the ($E$, $L_x$, $L_y$, $L_z$) space. We compare the results with the corresponding results using Friends-of-Friends. We know that in axisymmetrical potential of the MW, the orbital parameters $E$, $L$, and $L_z$ are known to be approximately conserved, whereas $L_x$ and $L_y$ evolve coherently \citep{Helmi00, knebe05, klement10, gomez10a, maffione15}. In a more realistic scenario, such as in our model which includes a live N-body disk and bulge, the conservation of these quantities are more strongly violated. However, after stars from a satellite interact with the stars in the MW and other satellites, some of them can still have similar $E$ and $\mathbf{L}$, due to very similar disruption history. This can be seen from the Aitoff projection of $\mathbf{L}$ shown in ($\theta$, $\phi$) in the left panel of Fig.1, where stars of the same origin tend to cluster. Similar clustering can also be seen in ($L_x$, $L_y$), ($L_x$, $L_z$), and ($L_z$, $E$) (see Fig. 2 a-c). Substructure identification methods can exploit these clusterings to identify star groups.

\subsection{\textsc{StarGO}}
\label{subsec:stargo}

\begin{table*}
\label{tab:stargo}

\centering
\caption{Results from \textsc{StarGO} applied to the mock catalogue in the ($E$, $L_x$, $L_y$, $L_z$) space}
\label{tab:stargo}
\begin{tabular}{|c|cccccccccc|}
\hline
i=0&&&&&&&&&& \\
GrpID& \textbf{A} & B & C & D &\textbf{E} & F & G&&& \\

 \hline
 Sat1 & & &&63&& 50&61&&&\\
 Sat2 & & &35&&& &&&&\\
 Sat9 & &26&&&&&&&& \\

 \hline
 $N_{\rm grp}$&\textbf{next}&48&39 & 63 &\textbf{next}& 50 & 66&&& \\
 purity(\%) &\textbf{iteration}& 54 & 90& 100&\textbf{iteration}&  100& 92 &&& \\
 \hline
 \hline
 i=1&\multicolumn{7}{c|} {A}&\multicolumn{3}{c|} {E}\\
 \cline{2-11}
 GrpID  &\textbf{A0} & A1 & A2 &\textbf{A3}& A4 & A5 & \multicolumn{1}{c|}{A6} & E0 & E1 & E2 \\
 \hline
 
Sat1 &&&&&&&\multicolumn{1}{c|}{}&180&35 & 52\\
Sat4 &&&45&&36&&\multicolumn{1}{c|}{}&&&\\
Sat5 &&&&&&&\multicolumn{1}{c|}{49}&&&\\
Sat7 &&27&&&&&\multicolumn{1}{c|}{}&&&\\
Sat9 &&&&&&49&\multicolumn{1}{c|}{}&&&\\

 \hline
 $N_{\rm grp}$&\textbf{next}&33&86&\textbf{next}&44&74 & \multicolumn{1}{c|}{64} & 182&35 & 52\\
 purity(\%) &\textbf{iteration}& 82 & 52&\textbf{iteration}& 82 &66 &\multicolumn{1}{c|}{77} & 99 & 100 & 100\\
 \hline
 \hline 
  i=2&\multicolumn{2}{c|}{A0}&\multicolumn{2}{c}{A3}&&&&&&\\
  \cline{2-11}
  GrpID  &\textbf{A0a}& \multicolumn{1}{c|}{A0b}&A3a&A3b&&&&&&\\
 \hline

 Sat5 &&\multicolumn{1}{c|}{}&&265&&&&&&\\
Sat7 &&\multicolumn{1}{c|}{143}&&&&&&&&\\
 \hline
 $N_{\rm grp}$&\textbf{next}&\multicolumn{1}{c|}{208}&1887&284&&&&&&\\
 purity(\%) &\textbf{iteration}& \multicolumn{1}{c|}{69} &spurious& 93&&&&&&\\
 \hline
 \hline
   i=3 &\multicolumn{3}{c}{A0a}&&&&&&&\\
 \cline{2-11}
 GrpID  &A0a0&\textbf{A0a1}&A0a2&&&&&&&\\
 \hline
  Sat4 &&&75&&&&&&&\\
  Sat7 &122&&&&&&&&&\\
 \hline
 $N_{\rm grp}$&131&\textbf{next}&80&&&&&&&\\
 purity(\%) & 93 &\textbf{iteration}&94&&&&&&&\\
 \hline
 \hline 
   i=4&\multicolumn{3}{c}{A0a1}&&&&&&&\\
\cline{2-11}
 GrpID  &\textbf{A0a1a}&A0a1b&A0a1c&&&&&&& \\
 \hline
  Sat7 &&92&30&&&&&&&\\
 \hline
 $N_{\rm grp}$&\textbf{next}&101&31&&&&&&&\\
 purity(\%) &\textbf{iteration}&91&97&&&&&&&\\
 \hline
  \hline
 i=5&\multicolumn{3}{c}{A0a1a}&&&&&&& \\
 \cline{2-11}
 GrpID  &A0a1a0&\textbf{A0a1a1}&A0a1a2&&&&&&&\\
 \hline
  Sat7 &113&&87&&&&&&&\\
 \hline
 $N_{\rm grp}$&138&\textbf{next}&117&&&&&&&\\
 purity(\%) &82&\textbf{iteration}&74&&&&&&&\\
 \hline
 \hline 
i=6&\multicolumn{2}{c}{A0a1a1}&&&&&&&&\\
 \cline{2-11}
 GrpID  &A0a1a1a&A0a1a1b&&&&&&&&\\
 \hline
  Sat4 &185&29&&&&&&&&\\
 \hline
 $N_{\rm grp}$&252&38&&&&&&&&\\
 purity(\%) &73&76&&&&&&&&\\
 \hline

	\end{tabular}
  \medskip
  
For each iteration, we list the identified satellites and the number of stars in the corresponding identified groups. $N_{\rm grp}$ denotes the total number of stars in each identified group.

\end{table*}

Following steps 1 -- 6 of the workflow (see Fig.~\ref{fig:workflow}), we apply SOM to the mock catalogue in the ($E$, $L_x$, $L_y$, $L_z$) space. Fig.~\ref{fig:stargo}Ia shows the training results after the application of SOM (step 1 of the algorithm) on the 2-D neural map. Each BMU is represented with the color and symbol according to the satellite of the associated stars, where the darker symbols represent multiple stars mapped to the same neuron. Some neurons are BMUs of stars belonging to different satellites, which can be seen with overlaid symbols. After we perform the group identification algorithm (steps 2--4), the neurons in seed groups are marked by blue pixels and the neurons with $u < u_{\rm thr}$ are marked by salmon pixels in Fig.~\ref{fig:stargo}Ib, same as Fig.~\ref{fig:u}c--d. Fig.~\ref{fig:stargo}Ic shows the identified star groups (Group A--G) using the same color coding as Fig.~\ref{fig:stargo}Ia. The group identification applied to individual groups requires six more iterations for Group A and one more iteration for Group E before we reach the end of the workflow  (see Fig.~\ref{fig:workflow}). For illustration, we show the detailed group identification procedure for two more iterations for Group A in Fig.~\ref{fig:stargo}II--III. Figure.~\ref{fig:treemap} shows the full schematic diagram for the hierarchical group identification from \textsc{StarGO} with the detailed results listed in Table~\ref{tab:stargo}.

We find that for most identified groups, the major contribution is from a single satellite. If the purity for a group is $\geq 60\%$, we identify the group with the corresponding satellite, which we refer to as the dominant contributor. Using this criteria, \textsc{StarGO} is able to find a total of 24 star groups, out of which 21 can be identified with satellites. One group is considered as a spurious group (Group A3a), which has roughly equal fraction of stars from sat4 and sat7 with purity $\sim 40\%$. The remaining two groups (Group B and  A2) have purity ranging from 50\%--54\% and thus cannot be strictly identified with a satellite using our criteria.
We find that all of the six major satellites (sat1, sat2, sat4, sat5, sat7, and sat9) in the mock catalogue can be identified with at least one group. The number fraction $f$ of stars of a satellite in the extracted volume that are identified with its corresponding groups is $\geq 5\%$ for all major satellites except sat2 ($f$=1.7\%). This is likely due to the fact that sat2 is more heavily disrupted compared to the other five major satellites. On the other hand, for the two largest contributors to the mock catalogue, sat 4 and sat 7, \textsc{StarGO} is able to identify 6.6\% (4 groups) and 12\% (7 groups) of the stars, respectively. Overall, \textsc{StarGO} is able to identify a total of 1850 stars from satellites within the analyzed volume that are the major contributors to the identified groups. This constitutes a fraction $f_{\rm tot}=10\%$ of the total number of stars in the mock catalogue.

\subsection{Friends-of-Friends}
\label{subsec:fof}

A widely used method of substructure identification is Friends-of-Friends, which is the standard procedure of finding halo groups used in cosmological simulations. In this case, the linking length $l_p$ is a key parameter, which is chosen empirically. The typical value of $l_p$ is set to be 0.2 times the inter-particle distance, which is a characteristic length scale used in the definition of dark matter halo. When FoF is applied to substructure identification in the integral-of-motion space, all ``distances" have units of angular momentum or energy, such that inter particle distances lose their physical meaning. Following \citep{Helmi00}, we set the characteristic length scale to be the dispersion of the total angular momentum of stars $\Delta L_{\mathrm{cluster}}$. We note that, similar to \textsc{StarGO}, we use a normalized  input space which is dimensionless. Thus, the dispersion of the angular momentum can be used as a scale  for every dimension. The linking length $l_p$ is set to be $\eta\Delta L_{\mathrm{cluster}}$, where $\eta$ is empirically chosen from a range of $\sim$ 0.1 -- 0.2. Finding optimal values of $\eta$ and determining $\Delta L_{\mathrm{cluster}}$ are problematic for substructure identification.

\begin{figure*}[tb]
\centering
\includegraphics[width=\linewidth]{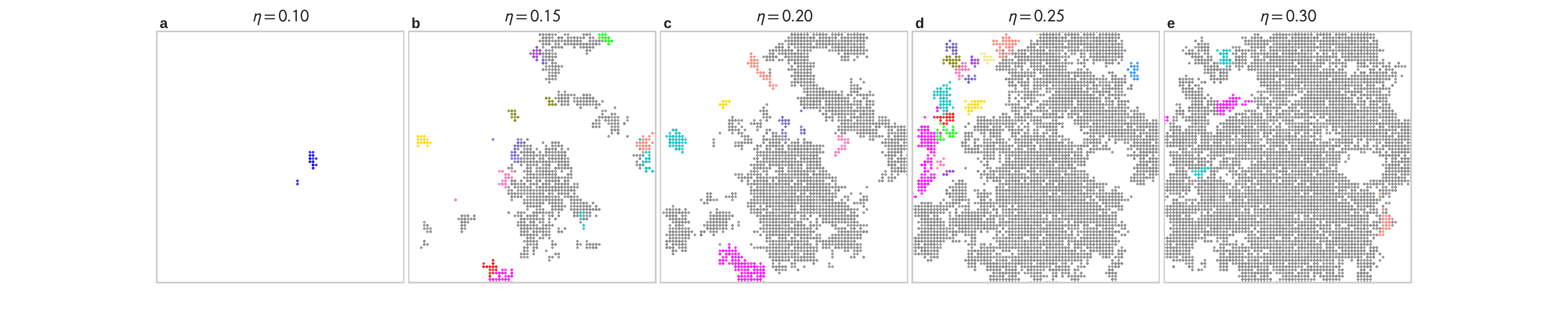}

\caption{Results from FoF applied to the mock catalogue in the ($E$, $L_x$, $L_y$, $L_z$) space, visualized on the 2-D neural map resulting from the initial application of SOM shown in Fig.\ref{fig:stargo}Ia. Star groups identified by FoF are shown in distinct colors using different values of $\eta$: a) $\eta=0.1$ b) $\eta=0.15$ c) $\eta=0.2$ d) $\eta=0.25$ e) $\eta=0.3$. The spurious group which cannot be identified with any satellite is colored in grey. }
\label{fig:fof}
\end{figure*}

Following the procedures from \citet{Helmi00}, we test the performance of FoF applied to the mock catalogue. We estimate $\Delta L_{\mathrm{cluster}}$ from the distribution of $L_z$. Specifically, we set it to be half of 68\% confidence interval, which roughly corresponds to the $\pm 1\sigma$ range for a normal distribution. We apply FoF using different values of $\eta$. In order to visualize the results from FoF for easy comparisons with \textsc{StarGO}, we again use the 2-D neural map. To do this, we map each star to its BMU resulting from the initial application of SOM. We use distinct colors to plot the stars of each group identified by FoF. As in the case of \textsc{StarGO}, we only consider groups with more than 30 stars. Fig.\ref{fig:fof}a--e shows the results for three different values of $\eta$. As we can see, the group identification is sensitive to $\eta$. The optimal results are found for $\eta=0.15$--0.25, which gives the maximum number of identified satellites within the extracted heliocentric volume and highest $f_{\rm tot}$. The results for $\eta=0.15$, $0.20$, and $0.25$ are listed in Tab.~\ref{tab:comp}.

For $\eta=0.20$, FoF is able to identify sat1, sat4, sat5, sat7, and sat9 with $f_{\rm tot}=3.9\%$, where 6 out of 7 groups can be identified with satellites. When $\eta$ is reduced to $0.15$, FoF is still able to identify  sat1, sat4, sat5, sat7 with $f_{\rm tot}=2.1\%$, but is unable to identify sat9. In this case 9 out of the 11 groups can be identified with satellites. On the other hand, for  $\eta=0.25$, FoF can identify sat1, sat2, sat4, sat9 with $f_{\rm tot}=4.5\%$, where 12 out of 13 groups can be identified with satellites. In contrast, \textsc{StarGO} is able to identify all the satellites with $f_{\rm tot}=10\%$. Interestingly, for all the three values of $\eta$, the group that cannot be identified with any single satellite (marked by grey pixels in Fig. 7) is the largest group, with roughly equal contributions from sat4 and sat7. This is likely due to the fact that sat4 overlaps heavily with sat7 (see Fig.~\ref{fig:proj_L} and Fig.~\ref{fig:proj_EL}). It results in weak clustering features such that FoF is barely able to distinguish them. This can clearly seen from Tab.~\ref{tab:comp}, where FoF gives very low values of $f$ for sat4 and sat7 for $\eta=0.20$ and is unable to identify sat7 for $\eta=0.25$. $\eta=0.15$ gives highest values of $f=2.1\%,1.8\%$ for sat4 and sat7 respectively, which are still below $f=6.6\%, 12\%$ obtained from \textsc{StarGO}. Similarly, FoF also fails to identify sat2 for $\eta=0.15$ and $\eta=0.2$, and gives low value of $f=1.4\%$ for $\eta=0.25$, whereas \textsc{StarGO} gives slightly better result of $f=1.7\%$. As mentioned before, this is likely due to the fact that sat2 has gone through severe disruption which results in weak clustering signal in the input space. Even for the optimal range of values of $\eta$, the variation of FoF results can be seen from the fact that sat5 can be easily identified for $\eta=0.15$ and $\eta=0.20$ but cannot be identified at all for $\eta=0.25$. On the other hand, \textsc{StarGO} gives higher value of $f = 16\%$ compared to $f = 14\%$ from the best case of FoF with $\eta=0.2$. Similarly, the identified fraction of stars from sat1 increases sharply from 3.7\% to 20\% as $\eta$ in increased from 0.20 to 0.25. Compared to such variations, \textsc{StarGO} is able to identify sat1 with a moderate value of $f=12\%$. 

For values of $\eta$ outside of the optimal range, FoF identifies fewer satellites or has even  lower values of $f$ and $f_{\rm tot}$ within the analyzed volume (shown in Fig.~\ref{fig:fof}). For $\eta=0.1$, FoF can find only one group of 36 stars ($f_{\rm tot}=0.19\%$), which is associated with sat7. For $\eta=0.3$, there are three groups identified from sat1, sat4 and sat9 with $f_{\rm tot}=1.5\%$.

\begin{table*}
\centering
\caption{Results from FoF applied to the mock catalogue in the ($E$, $L_x$, $L_y$, $L_z$) space}
\label{tab:fof}
	\begin{tabular}{|cccccccccccccc|}
		\hline
		\multicolumn{14}{|c|}{$\eta=0.15$}\\
		\hline

  GrpID  & A & B & C & D & E & F & G & H & I & J & K &&\\

 \hline
 Sat1 &&&&&&41&&&&&&& \\
 Sat4 &&&&&29&&&35&22&33 &&& \\
 Sat5 &&102&&&&&&&&&32 && \\
 Sat7 &&&47&42&&&&&&&&&\\

 \hline
 $N_{\rm grp}$&1919&102&50&44&42&41&40&35&34&33&32&&\\
 purity(\%) & spurious & 100 & 94 & 95 & 69 & 100& spurious & 100 & 65 & 100 & 100 &&\\
 \hline
\hline	
\multicolumn{14}{|c|}{$\eta=0.20$}\\
		\hline

  GrpID  & A & B & C & D & E & F & G&&&&&& \\

 \hline
 Sat1 &&& 99 &&&36&&&&&&& \\
 Sat4 &&&&&40&&&&&&&&  \\
 Sat5 &&282&&&&&&&&&&& \\
 Sat7 &&&&&&&30&&&&&&\\
 Sat9 &&&&52&&&&&&&&&  \\

 \hline
 $N_{\rm grp}$&5276&302&100&57&40&36&30&&&&&&\\
 purity(\%) & spurious & 93 & 99 & 91 &100 & 100 & 100&&&&&&\\
 \hline
	
	\hline
	\multicolumn{14}{|c|}{$\eta=0.25$}\\
			\hline

  GrpID  & A & B & C & D & E & F & G & H & I & J & K & L & M\\

 \hline
 Sat1 &  &272&95& &72&67&57&43&40&36&31&& \\
  Sat2 &  &&& &&&&&&&&28& \\

  Sat4 &  &&& &&&&&&&&&31 \\
 Sat9 &  & & &67& & && & &  &&&\\

 \hline
 $N_{\rm grp}$&9277&274 & 95& 74 & 72 & 67 & 57&43 & 40 &36 & 31&31&\\
 purity(\%) & spurious & 99& 100& 90& 100& 100& 100&100& 100& 100& 100&90 &100\\
 \hline
	\end{tabular}
  \medskip

\end{table*}

\begin{table*}
\centering
\caption{Comparison of $f$ between \textsc{StarGO} and FoF}
\label{tab:comp}
	\begin{tabular}{|cccccccccccc|}
		\hline
 Satellite  && sat1 & sat2 & sat3 & sat4 & sat5 & sat6 & sat7& sat8 & sat9 & Total \\

 \hline
$N_{\mathrm{sat}}$&&3609&1970&59&5598&1961&9&4985&38&341 & 18570 \\
\hline
$f$(\%) &\textsc{StarGO}&12  &1.7&& 6.6&16&&12&&14 &10 \\
            &FoF ($\eta=0.15$)&1.1&&&2.1 &6.8&&1.8&&&2.1\\
            &FoF ($\eta=0.2$)&3.7&&&0.1 &14&&0.6&&15&3.9\\
            &FoF ($\eta=0.25$)&20  &1.4&&0.5& &&&&22&4.5\\

 \hline

	\end{tabular}
  \medskip

\end{table*}

\section{Conclusion} \label{sec:sum}

In this paper we present a new substructure identification method \textsc{StarGO} that identifies and visualizes star groups hierarchically on top of a 2-D neuron map. Our algorithm first maps the multidimensional phase space coordinates of stars into a 2-D map using SOM while conserving the topological structure of the dataset. It then identifies a hierarchy of star groups adaptively according to the significance of clustering at each step. 

We test our algorithm using a mock catalogue of stars within a heliocentric radius of 10 kpc generated from a simulated MW-like system, and compare the results against that from an FoF algorithm. In the tests we take into account observational errors that are expected for K-giants in the Gaia DR2 and LAMOST DR5 catalogues. In comparison to FoF, \textsc{StarGO} is able to identify star groups dominated by each of the six major satellites, whereas FoF is able to identify at most five even after optimizing the linking length. In addition, \textsc{StarGO} can identify a higher fraction of stars from almost all the satellites compared to FoF (see Table.~\ref{tab:comp}). If we consider the number of stars from the dominant satellite in each group, we find that \textsc{StarGO} is able to identify a total of 10\% of the total satellite population in the extracted heliocentric volume, whereas for FoF this fraction is below 4.5\%. 

In conclusion, \textsc{StarGO} is able to identify star groups efficiently by combining the sensitivity and visualization ability of SOM with an adaptive clustering algorithm.
The adaptive group identification procedure allows us to systematically search for substructures while avoiding uncertainties from nuisance parameters. Overall, the results from \textsc{StarGO} are better than the results from FoF, even when using an optimized linking length. \textsc{StarGO} is an ideal tool to explore high dimensional data set from the recently released Gaia DR2. Our method will be particularly useful for studies of the inner stellar halo, for example when applied to the cross-match of Gaia DR2 and spectroscopic surveys.

\section*{Acknowledgements}
Z.Y. gratefully acknowledges Jingying Lin for inspiring discussions about SOM.  Z.Y. thanks Xiang Xiang Xue for sharing her expertise in applying FoF to the real data, and Chao Liu for insightful discussions about the algorithm of \textsc{StarGO}. Z.Y. is also indebted to Yi Peng Jing, Yu Luo, and Hong Guo for commenting on the early draft of this paper. All the authors thank the anonymous referee for valuable and constructive comments which greatly improved this work. This work is supported by the National Key Basic Research Program of China (No. 2015CB857003). Z.Y. and P.B. acknowledge the support of NSFC-11533006. J.C. and X.K. acknowledge the support of NSFC-11333008. J.X.H. acknowledge the support of JSPS Grant-in-Aid for Scientific Research JP17K14271. M.C.S. acknowledges fiancial support from the CAS One Hundred Talent Fund and from NSFC grants 11673083 and 11333003. This work was also supported by the National Key Basic Research Program of China 2014CB845700. We gratefully acknowledge the use of the High Performance Computing Resource in the Core Facility for Advanced Research Computing at Shanghai Astronomical Observatory.

\bibliography{ms}
\bibliographystyle{apj}

\end{document}